\documentclass[conference]{IEEEtran}
\usepackage[numbers,sort&compress]{natbib}

\newcommand{\code}[1]{\texttt{#1}}
\usepackage{listings}
\usepackage[svgnames]{xcolor}
\definecolor{mycolor}{HTML}{FF0013}
\usepackage[]{mdframed}
\usepackage[colorlinks = true,
            linkcolor = blue,
            urlcolor  = blue,
            citecolor = blue,
            anchorcolor = blue]{hyperref}

\mdfdefinestyle{tag}{%
    roundcorner=5pt,
    innerlinewidth=1pt, innerlinecolor=RubineRed,%
    middlelinewidth=2pt, middlelinecolor=white,%
    outerlinewidth=1pt, outerlinecolor=Melon,%
    backgroundcolor=black!10,
    innerleftmargin=5pt, innerrightmargin=5pt, innertopmargin=5pt, innerbottommargin=5pt%
}

\definecolor{eclipseStrings}{RGB}{42,0.0,255}
\definecolor{eclipseKeywords}{RGB}{127,0,85}
\colorlet{numb}{magenta!60!black}

\lstdefinelanguage{json}{
    basicstyle=\fontsize{8}{10}\selectfont\ttfamily,%\normalfont\ttfamily,
    commentstyle=\color{eclipseStrings}, % style of comment
    stringstyle=\color{eclipseKeywords}, % style of strings
    showstringspaces=false,
    showlines=false,
    breaklines=true,
    frame=lines,
    frame=,
    string=[s]{"}{"},
    comment=[l]{:\ "},
    morecomment=[l]{:"},
    literate=
        *{0}{{{\color{numb}0}}}{1}
         {1}{{{\color{numb}1}}}{1}
         {2}{{{\color{numb}2}}}{1}
         {3}{{{\color{numb}3}}}{1}
         {4}{{{\color{numb}4}}}{1}
         {5}{{{\color{numb}5}}}{1}
         {6}{{{\color{numb}6}}}{1}
         {7}{{{\color{numb}7}}}{1}
         {8}{{{\color{numb}8}}}{1}
         {9}{{{\color{numb}9}}}{1}
}

\definecolor{codegreen}{rgb}{0,0.6,0}
\definecolor{codegray}{rgb}{0.5,0.5,0.5}
\definecolor{codepurple}{rgb}{0.58,0,0.82}
\definecolor{backcolour}{rgb}{0.95,0.95,0.92}

\lstdefinestyle{mystyle}{
 backgroundcolor=\color{backcolour},   
 commentstyle=\color{codegreen},
 keywordstyle=\color{magenta},
 numberstyle=\tiny\color{codegray},
 stringstyle=\color{codepurple},
 basicstyle=\ttfamily\small,
 breaklines=true,
 breakatwhitespace=false, % or true, if you prefer
 breakindent=0pt, % Adjusts the indent of broken lines              
 captionpos=b,                    
 keepspaces=true,                 
 showspaces=false,                
 showstringspaces=false,
 showlines=false,
 showtabs=false,                  
 tabsize=2
}

\ifCLASSINFOpdf
   \usepackage[pdftex]{graphicx}

\else
   \usepackage[dvips]{graphicx}
\fi

\begin{document}

\title{Accelerating Radio Spectrum Regulation\\ Workflows with Large Language Models (LLMs)}

\author{\IEEEauthorblockN{Amir Ghasemi, Paul Guinand}
\IEEEauthorblockA{Communications Research Centre (CRC) Canada\\
%Innovation, Science and Economic Development Canada\\
Ottawa, Ontario, Canada\\
Email: \{amir.ghasemi, paul.guinand\}@ised-isde.gc.ca}}
%\and
%\IEEEauthorblockN{Homer Simpson}
%\IEEEauthorblockA{Twentieth Century Fox\\

\maketitle

\begin{abstract}

Wireless spectrum regulation is a complex and demanding process due to the rapid pace of technological progress, increasing demand for spectrum, and a multitude of stakeholders with potentially conflicting interests, alongside significant economic implications. To navigate this, regulators must engage effectively with all parties, keep pace with global technology trends, conduct technical evaluations, issue licenses in a timely manner, and comply with various legal and policy frameworks.

In light of these challenges, this paper demonstrates example applications of Large Language Models (LLMs) to expedite spectrum regulatory processes. We explore various roles that LLMs can play in this context while identifying some of the challenges to address. The paper also offers practical case studies and insights, with appropriate experiments, highlighting the transformative potential of LLMs in spectrum management.

\end{abstract}

\begin{IEEEkeywords} large language models, artificial intelligence, spectrum management, spectrum regulation, generative AI
\end{IEEEkeywords}

\IEEEpeerreviewmaketitle

\section{Introduction}

Wireless spectrum regulation is complex and challenging given the rapid pace of technological progress, increasing demand for spectrum, and a wide range of stakeholders with competing interests. Regulatory processes often involve labour-intensive tasks such as reviewing technical submissions, coordinating frequency assignments among various users, consulting with industry stakeholders, enforcing compliance with regulations, and updating policy frameworks to reflect the latest technological advancements and market needs. In this paper we investigate the possible use of Generative Artificial Intelligence (GenAI) to expedite these processes. 

 GenAI is a branch of AI that involves generating new outputs based on one or more inputs. Both inputs and outputs can take various formats such as text, image, audio, video, or other forms of data. Within this context, text-to-text models where a text input (prompt) is used to produce a text output have various applications such as document summarization, question-answering, language translation, and computer code generation, to name a few. Techniques developed for these tasks are collectively referred to as natural language processing (NLP), within which, LLMs form a subset.

%Initially, NLP models were relatively simple and rule-based. These models relied on a set of hand-coded rules to interpret and generate language which was quite limited due to the complexity and variability of human language. A breakthrough came with the introduction of machine learning, and especially deep learning algorithms which allowed models to learn and make predictions based on the data they were fed as opposed to relying on hand-coded rules. The development of recurrent neural networks (RNNs) and particularly Long Short-Term Memory networks (LSTMs) enabled models to remember longer strings of text, making them better at understanding context.

%However, it wasn't until the introduction of Transformer architecture in 2017 with the paper "Attention is All You Need" \cite{attention} that NLP took a significant leap forward. The Transformer model abandoned the sequential nature of RNNs and LSTMs for a parallel processing approach that allowed for much faster training times and better handling of long-range dependencies in text.

Today, transformers \cite{attention} form the building blocks of LLMs, which are distinguished by their size (some containing hundreds of billions of parameters or more) and their ability to perform a wide range of language tasks without task-specific training. They can write essays, summarize text, translate languages, and even generate code, demonstrating a level of linguistic capability and versatility that was hard to imagine a decade ago.

The remainder of this paper is organized as follows: Key applications of LLMs in spectrum regulation domain as well as some of their main challenges and limitations in this context are presented in Sections II and III, respectively. Instances of LLMs' practical applications in regulatory tasks are demonstrated in Section IV, followed by a summary of key practical lessons and considerations in Section V. Finally, Section VI provides some concluding remarks.  

\section{Impact of LLMs on Radio Spectrum Regulation}

This section outlines a number of regulatory tasks that can be improved with LLMs. We will discuss how LLMs can streamline and accelerate processes, aid in decision-making, and ensure more holistic responses to regulatory inquiries.

\subsection{Stakeholder Consultations}

 Preparing radio spectrum regulatory consultations typically involves conducting an initial research phase as well as having an understanding of legal frameworks, diverse policy implications, and industry-specific technicalities. LLMs can expedite the research phase by summarizing relevant international efforts (e.g., those of ITU or other regional bodies) or approaches taken by other national regulators within their jurisdictions. They can also provide a summary of the feedback and comments received from various types of stakeholders in prior or similar consultations to help regulators better anticipate and understand potential concerns. LLMs can also help in automating the drafting process via providing initial templates, suggesting language based on past successful consultations, and ensuring compliance with legal and procedural norms. This not only saves time but also reduces the potential for human error. Furthermore, LLMs can assist with processing and distilling comments received through the consultation process to speed up the final decision-making.
 
\subsection{Rules as Code}

Radio spectrum regulations such as radiated power and antenna height limits could vary based on frequency, geographic location, application, and existence of multiple licensees. These rules are typically embedded within regulatory documents and are often wrapped in legal and technical language, making it difficult for end-users to understand and comply. The Rules as Code (RaC) concept aims to address this difficulty by converting legislative rules, regulations, and policies embedded within both structured and unstructured data sources into machine-readable code \cite{RaC, RaC2}. Thus, RaC represents a significant shift towards integrating technology with legislative processes to ensure that regulations can be more effectively administered, interpreted, and applied in a consistent and accessible manner.

LLMs offer a one-stop solution for RaC as they are 1) capable of understanding complex legal and regulatory documents and databases, 2) can identify and extract rules, conditions, and constraints that would be applicable under different scenarios from these sources \cite{LLM_RaC}, 3) can directly codify these rules into various programming languages thanks to their code generation capabilities, and 4) can generate test cases to validate the code to ensure it accurately reflects the original regulatory intent.

\subsection{Knowledge-Base Question Answering}

One of the major issues faced by spectrum regulators is the time-consuming task of sifting through many sources of information such as policy and regulatory documents, technical specifications, and license conditions, in order to be able to answer regulatory inquiries. LLMs can process and analyze large amounts of data rapidly, providing quick responses to queries. Automating the question-answering process with LLMs can reduce the need for extensive human labor, cutting down on the time and cost associated with regulatory inquiries.

In addition to saving time and labor, by drawing on authoritative knowledge bases, LLMs can provide precise and more holistic responses to complex regulatory queries requiring analysis of multiple factors and data sources. Furthermore, LLMs make information more accessible to non-experts by translating technical regulatory language into more understandable terms, thus enhancing the capability of less experienced staff to handle more complex problems efficiently. This has the added benefit of democratizing access to information for the general public. 

A practical implementation of such question-answering system will be detailed in Section \ref{sec:QA}.

\subsection{Automating Processes via LLM Agents}

LLMs have the potential to revolutionize radio spectrum regulatory processes by functioning as autonomous agents capable of executing tasks efficiently and accurately. While LLMs are not inherently agents, as they do not have any goals, they can be prompted to act like one \cite{llm_agent1, llm_agent2}. These LLM-powered agents can be instrumental in streamlining administrative procedures, particularly in contexts such as license issuance and renewal, coexistence studies, and interference complaint investigations among other things. This type of advanced application is enabled by LLM agents' \textit{reasoning} capability where they can break down complex requirements into a series of individual tasks, choose the right tool for each task (e.g, run a query against a database, search the Internet for latest industry developments, or run a radio propagation simulator), and eventually interpret, summarize, and present the final output. 

\subsubsection{Training LLMs as Regulatory Agents}

LLMs can be trained to act as regulatory agents by leveraging their natural language understanding and generation capabilities. Through appropriate training and fine-tuning, these models can develop an understanding of the intricate rules, regulations, and procedures governing radio spectrum management. They can process complex legal documents, identify relevant clauses, and interpret them in the context of specific regulatory tasks, query various databases, run radio network or coexistence simulations via application-specific programming interfaces (APIs) and even interpret the final results to make recommendations. This enables LLMs to act as knowledgeable agents complementing the human expertise, capable of comprehending regulatory requirements and making informed choices autonomously.

\subsubsection{Streamlining Administrative Processes}

LLM-powered agents hold significant promise in streamlining administrative processes, particularly those related to license issuance and renewal. These tasks often involve extensive paperwork, data verification, and adherence to regulatory compliance. LLM agents can efficiently handle data entry, document verification, and even communication with applicants or license holders. By automating these routine, yet time-consuming tasks, LLMs can reduce the administrative burden on regulatory bodies, leading to quicker and more efficient processes. This can result in faster response times, reduced paperwork, and minimized errors in the regulatory workflow.

\subsubsection{Importance of Human Oversight}

While LLM agents offer substantial advantages in terms of efficiency and accuracy, it is crucial to emphasize the importance of human oversight in LLM-agent-assisted processes. Regulatory decisions involving spectrum can have far-reaching consequences and human judgment remains essential in cases involving ambiguity, novel applications, or fairness considerations. Human oversight ensures that the decisions made by LLM agents align with regulatory intent, legal precedents, and ethical guidelines. Furthermore, it provides a mechanism for addressing exceptional cases that may fall outside the scope of routine procedures. Because of these concerns various jurisdictions are developing guidelines on the use of AI in decision making processes (see e.g., \cite{automated_decision_tbs}).

Human oversight also plays a critical role in monitoring LLM agents' performance, ensuring that they continue to operate effectively and without bias. Regular audits and reviews by human experts help identify and rectify any potential issues or biases that may arise in the LLM's decision-making process. This synergy between LLM agents and human oversight strikes a balance between efficiency and regulatory compliance, ultimately leading to more reliable and fair outcomes.

\section{Challenges and Ethical Considerations}

Leveraging LLMs in the government regulatory and policy-making domain holds significant promise for enhancing efficiency of the processes. However, it also comes with various challenges and limitations which should be taken into consideration. In what follows, we will provide a high-level overview of these challenges.

\subsection{Unconscious Bias}

GenAI may amplify existing unconscious biases in the data (e.g., those related to ethnicity, gender, or other demographic aspects), leading to discriminatory or non-representative content, potentially harming diversity in ideas and perspectives \cite{llm_bias}. To address these, we need to test for biases in the training data, as well as model outputs both before deploying a system, and on an ongoing basis. The process of improving LLMs' generated output to make it more consistent with a desired set of values is typically referred to as \textit{alignment}. A recent survey of alignment techniques such as Reinforcement Learning from Human Feedback (RLHF) and Direct Preference Optimization (DPO), where the reward inferred from human preferences is used to steer LLMs toward generating better outputs, can be found in \cite{alignment}. 

\subsection{Inaccuracy}

Output of generative models such as LLMs can be misleading, appearing credible but being factually wrong (so-called \textit{hallucination}) \cite{hallucinate}, or outdated. This in turn could misinform policies or public communication and lead to loss of public/stakeholders trust. Therefore, it is important to keep LLMs grounded in facts by providing them access to authoritative sources of domain knowledge and engineering prompts that would discourage hallucinations. An example of this will be demonstrated later in Section \ref{sec:QA}.

\subsection{Automation Bias}

Relying too much on AI-generated outputs and decisions can lead to \textit{automation bias} where automatically generated outputs are preferred even when faced with contradictory evidence suggesting other manually-derived outputs might be better \cite{automation_bias}. This can potentially limit creativity and critical thinking of the domain experts. To address this issue, it is essential for users to critically assess whether they need generative AI for a task, be aware of the strengths and weaknesses of this technology, and to use generative AI as a complement to human judgment instead of a substitute.

\subsection{Legal Risks}

Using generative AI such as LLMs in the regulatory and policy-making domains poses risks to privacy, intellectual property rights, and procedural fairness due to the data and processes involved. For instance, code that is automatically generated by an LLM might be a reproduction of existing code that is not licensed for commercial use. This calls for a thorough risk analysis of the training data used to train off-the-shelf LLMs and/or carefully choosing data sources used for training or fine-tuning internal models.

\section{Case Studies and Real-World Applications}
\label{sec:case_study}

In this section we give a few specific examples of real-world experimentation with LLMs to transform some of the radio spectrum regulation processes. 

\subsection{Rules as Code and Knowledge Graphs}

In this section we illustrate an example of machine-readable rules extraction and knowledge graph creation using an LLM. This can be achieved in a multitude of ways by e.g., providing the related regulatory document(s) to a local or API-based LLM, or allowing an LLM agent to search the web to find pertinent document(s) for rule extraction. For illustration, we employ the latter approach using ChatGPT-4 interface to extract technical rules related to operation in Personal Communication Systems (PCS) bands in Canada. The prompt and response are shown in the following where the ChatGPT-4 has found the related document \cite{srsp510} and then correctly extracted and structured the radiated power and antenna height limits under different scenarios in machine-readable JavaScript Object Notation (JSON) format.

\begin{figure}[!t]
\centering
\includegraphics[width=3.45in]{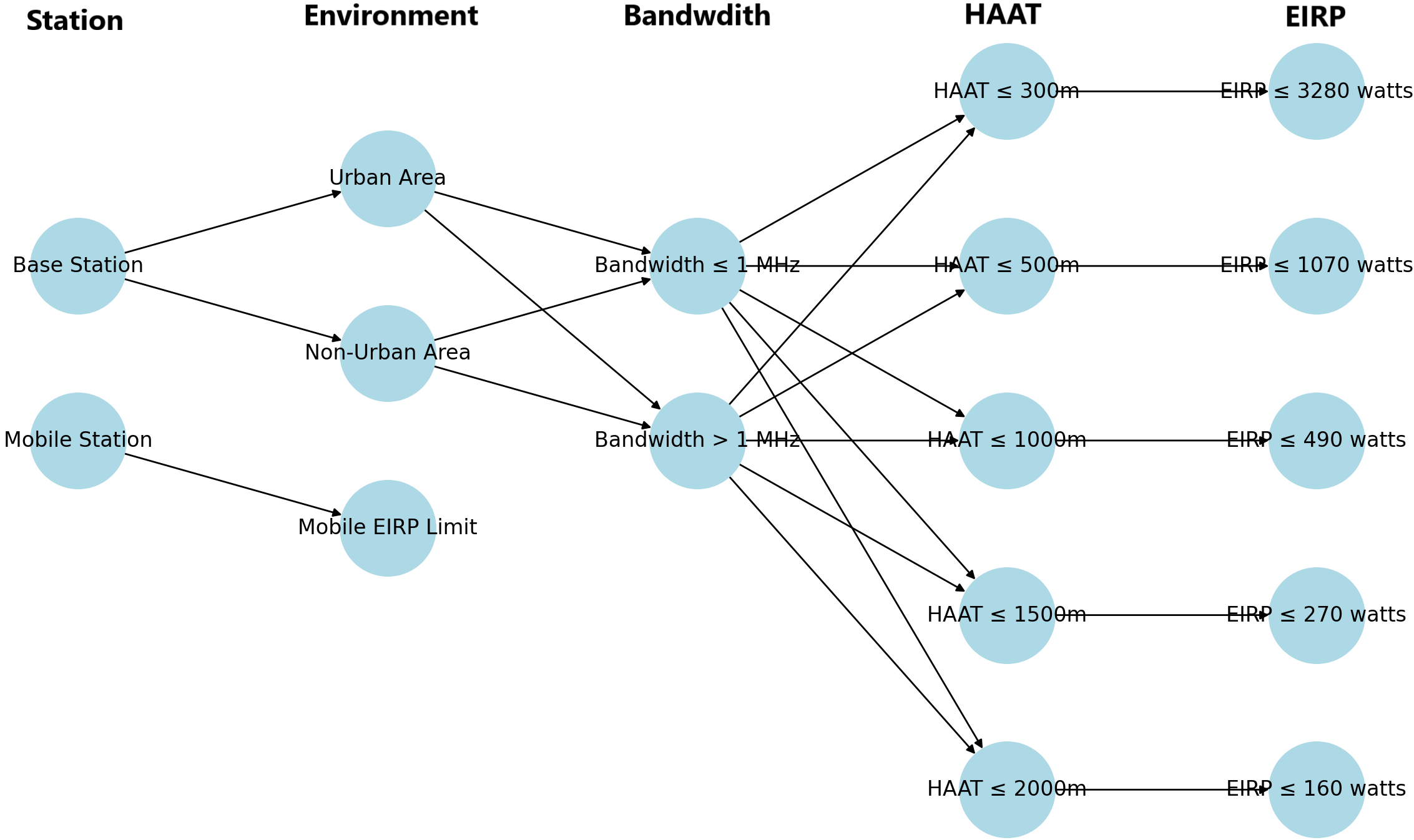}
\caption{Knowledge graph of technical rules for PCS bands in Canada, created by GPT-4.}
\label{fig_pcs_canada_kg}
\end{figure}

\begin{center}
\textbf{PCS Band Technical Parameters Extraction}
\end{center}
\vspace{1pt}
\begin{mdframed}[style=tag, backgroundcolor=black!5, linewidth = 1.5, innertopmargin=5pt, innerbottommargin=0pt]
%\hline
%\hline
\vspace{3pt}
\noindent
\textbf{Prompt:}
% \begin{minted}[breaklines,fontsize=\normalsize,breaksymbolleft=]{text}
%   What are the radiated power and antenna height limits in the PCS bands in Canada? Return the answer in JSON format.
% \end{minted}
\begin{lstlisting}[language=]
What are the radiated power and antenna height limits in the PCS bands in Canada? Return the answer in JSON format.
\end{lstlisting}
%\vspace{2pt}
\textbf{Response:}
%\vspace{4}
%\hline
%\vspace{4}
%\caption{PCS Band Extracted Technical Parameters in JSON Format}
\begin{lstlisting}[language=json]
 {"PCS Bands Canada": {
    "Base Stations Less Equal 1MHz": {
      "Max eirp": {
        "HAAT_up_to_300m": "3280 watts",
        "Urban_Areas": "1640 watts",
        "Height_Restrictions": [
          {"HAAT_up_to_500m": "1070 watts"},
          {"HAAT_up_to_1000m": "490 watts"},
          {"HAAT_up_to_1500m": "270 watts"},
          {"HAAT_up_to_2000m": "160 watts"}
        ]
      }
    },
    "Base_Stations_More_1MHz": {
      "Max_eirp_per_MHz": {
        "HAAT_up_to_300m": "3280 watts",
        "Urban_Areas": "1640 watts",
        "Height_Restrictions": [
          {"HAAT_up_to_500m": "1070 watts per MHz"},
          {"HAAT_up_to_1000m": "490 watts per MHz"},
          {"HAAT_up_to_1500m": "270 watts per MHz"},
          {"HAAT_up_to_2000m": "160 watts per MHz"}
        ]
      }
    },
    "Mobile_Stations": {
      "Max_eirp": "2 watts"
    }
  }
}
 \end{lstlisting}
 \end{mdframed}
\vspace{4pt}
The output of the LLM can further be used in downstream tasks such as creating a knowledge graph capturing the relationship between various entities such as frequencies, radio station type, antenna height above average terrain (HAAT), and equivalent isotropically-radiated power (EIRP). This process, too, can be automated via the LLM agent which, this time, leverages its code interpreter tool to plot the knowledge graph as shown next.
%\hline
%\hline
\vspace{4pt}
\begin{mdframed}[style=tag, backgroundcolor=black!5, linewidth = 1.5, innertopmargin=5pt, innerbottommargin=0pt]
\vspace{2pt}
\noindent
\textbf{Follow-up Prompt 1:}
%\begin{minted}[breaklines,fontsize=\normalsize,breaksymbolleft=]{text}
\begin{lstlisting}[language=]
 Extract a knowledge graph from these rules and return as an image.
\end{lstlisting}
%\end{minted}
\vspace{4pt}
\textbf{Response:}
Resulting image is shown in Fig. \ref{fig_pcs_canada_kg}.
\vspace{4pt}
%\hline
%\hline
\vspace{2pt}
\end{mdframed}

It is further possible to create an ontology based on the technical rules extracted from the documents and datasets. This is illustrated in the next example.
\begin{mdframed}[style=tag, backgroundcolor=black!5, linewidth = 1.5, innertopmargin=5pt, innerbottommargin=0pt]
\vspace{4pt}
\noindent
\textbf{Follow-up Prompt 2:}
%\begin{minted}[breaklines,fontsize=\normalsize,breaksymbolleft=]{text}
\begin{lstlisting}[language=]
Create an ontology based on the extracted information.
\end{lstlisting}
%\end{minted}
\vspace{4pt}
\textbf{Response:}
\vspace{5pt}
%\begin{minted}[breaklines,fontsize=\small,breaksymbolleft=]{text}
\begin{lstlisting}[language=]
Concepts:
- BaseStation
  - UrbanBaseStation
  - NonUrbanBaseStation
- MobileStation
- HAAT (Height Above Average Terrain)
- EIRP (Equivalent Isotropicaly Radiated Power)
- Bandwidth

Properties:
- hasEIRP: BaseStation -> EIRP
- hasHAAT: BaseStation -> HAAT
- hasBandwidth: BaseStation -> Bandwidth
- hasMaxEIRP: MobileStation -> EIRP

Relationships:
- UrbanBaseStation IS-A BaseStation
- NonUrbanBaseStation IS-A BaseStation

Constraints:
- EIRP and HAAT have numerical values associated with specific limits.
- Bandwidth is a characteristic that affects the EIRP.
- MobileStation has a maximum EIRP of 2 watts.
- UrbanBaseStation and NonUrbanBaseStation have different EIRP limits based on HAAT.
\end{lstlisting}
%\end{minted}
\vspace{0.5pt}    
\end{mdframed}

Distilling regulatory documents and datasets into structured knowledge graphs provides for better inter-connectivity among disparate data sources and enables enhanced retrieval of information thereby increasing the utility of information.

\subsection{RetrievalQA System}
\label{sec:QA}

Next, we showcase a retrieval-augmented generation (RAG) based question answering (QA) system \cite{RAG1, RAG2} capable of leveraging all of the public and internal regulatory documents of Canada's national spectrum regulator. This type of LLM application, which we will refer to as RetrievalQA hereafter, is of particular interest in the context of regulation and policy-making as it grounds the LLM in domain knowledge and authoritative data sources which helps it produce accurate, factual responses. 

Next, we will give an overview of the implemented proof-of-concept RetrievalQA system and highlight its potential benefits. In Section \ref{sec:lessons} we will share some of the lessons learned during the process.

The central concept in RAG systems is to ground the LLM in facts by providing it with relevant and accurate context and then prompt it to only draw from the provided context to respond using a prompt template such as the following:

\vspace{4pt}
\begin{mdframed}[style=tag, backgroundcolor=black!5, linewidth = 1.5, innertopmargin=5pt, innerbottommargin=0pt]
\vspace{2pt}
\noindent
\textbf{RAG Prompt Template:}
%\begin{minted}[breaklines,fontsize=\normalsize,breaksymbolleft=]{text}
\begin{lstlisting}[language=]
 You are a helpful, respectful and honest assistant. Use the following context information to answer the user's question. If you don't know the answer, just say that you don't know, don't try to make up an answer. Context: {context} Question: {question}
 \end{lstlisting}
%\end{minted}
%\vspace{4}
%\vspace{2}
\end{mdframed}

\noindent where \code{\{context\}} and \code{\{question\}} are variables determined at run-time.

In order to find and provide relevant context from a large set of documents, a retriever is implemented which typically uses semantic search (as opposed to keyword search). Since semantic meaning can often vary in different parts of longer documents, text is often split into smaller chunks which are then tokenized and converted to vector representations (vectors of floating numbers, referred to as embeddings) using pretrained models, and subsequently stored in a vector database. Such vector representation allows for effective semantic search where the user query is embedded using the same embedding model and is then compared with all text-chunk embeddings within the vector database. By returning a few of the most relevant document chunks, RAGs have enabled the use of LLMs within specialized domains without the need to fine-tune. 

\begin{figure}[!t]
\centering
\includegraphics[width=3.6in]{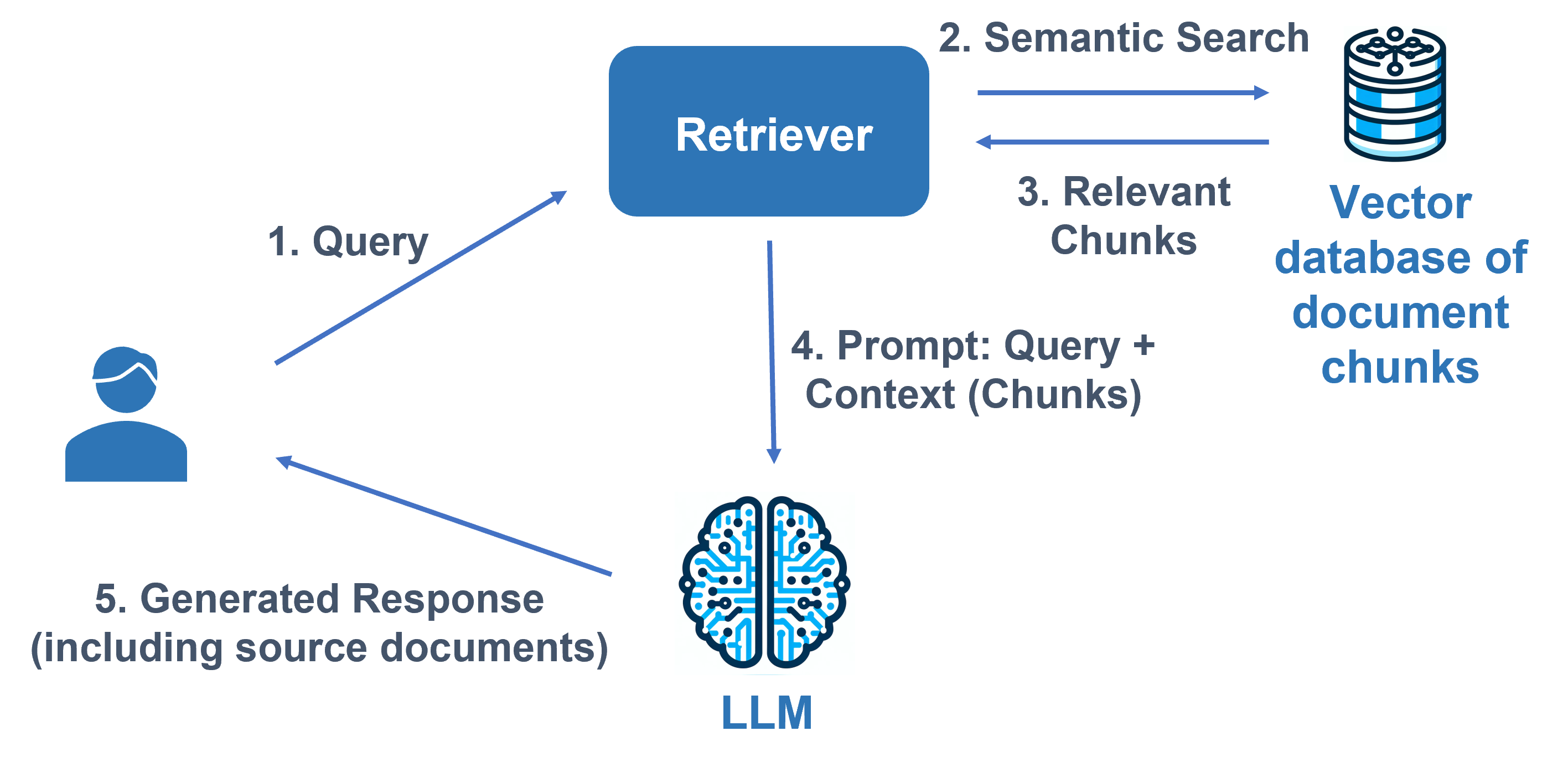}
\caption{Block diagram of the RAG-based question answering system.}
\label{fig_RAG}
\end{figure}

Fig. \ref{fig_RAG} shows a block diagram of our implemented RAG architecture. We employ the open-source frameworks, LangChain \cite{langchain} and llama.cpp \cite{llamacpp} for orchestration of the pipeline and retriever, and to interact with the LLM. We use the Faiss library \cite{faiss} as our vector database to store the vector representation of document chunks. We also leverage a locally-hosted version of the Mistral-7B LLM \cite{mistral7b} which comes with the Apache 2.0 license and can be used without restrictions. Despite being a relatively small 7-billion parameter model, Mistral has shown good performance in public benchmarks. The use of a locally-hosted LLM is of particular note as for reasons of privacy and data sovereignty it may be desirable to avoid using an external LLM. We use a virtual machine with a single NVIDIA V100 GPU to run the RAG.

For this experiment, the system is provided with approximately $2000$ documents in both PDF and HTML formats which are then parsed, chunked, and embedded. We then proceed with asking in-domain questions on a range of spectrum regulatory topics. While we omit sample queries and responses here due to limited space, our experience so far has shown great promise with the system being able to answer both high-level as well as very specific questions on a wide range of regulatory topics such as spectrum auction setup, technical rules, license conditions, and stakeholder sentiments in consultation responses. This initial success has generated a great deal of interest internally and serves as a foundation to build upon.

\subsection{Wireless License Application}

For many wireless applications, the user/operator needs to apply to the national spectrum regulator for a license before they can operate. The information required during the application as well as technical parameters and conditions associated with the license can vary widely depending on frequency, application, location, or time period which, in turn, renders the application process complex and error-prone, especially for individuals or smaller entities. This is an area where the power of LLMs could be leveraged to facilitate the application process by providing clear instructions to applicants.

Using OpenAI's "Create a GPT" framework, we have built a Wireless License Application chat-bot for Canada that guides applicants through the application process by providing clear instructions and helps to ensure higher-quality data entries. This has been achieved by providing the chat-bot with a document outlining information required for various license applications and assigning it the specific role of helping the applicants throughout the application process. The experimental chat-bot then converses with applicants and
%, accessible at \MYhref{https://chat.openai.com/g/g-lfnKWoAUN-wireless-license-application}{https://chat.openai.com/g/g-lfnKWoAUN-wireless-license-application}, 
converts the inputs into a machine-readable JSON format, ready for consumption by internal data bases of the regulator.

\section{Lessons Learned}
\label{sec:lessons}

In what follows, we will share some of the lessons learned during the implementation of our LLM-based question-answering system. 

\subsection{Retrieval Bottleneck}

Not surprisingly, the retrieval step can become a performance bottleneck for RAG systems as even the most advanced LLM can not produce good responses if the context (i.e., document chunks) provided by the retriever is not relevant, suitable, or accurate. A variety of techniques have been employed to address this issue. 

The process of document retrieval involves a balance between creating smaller documents that accurately reflect their content through embedding vectors and maintaining sufficiently large chunks to preserve context. Expanded context retrieval, aims to balance these competing needs by using small chunks during semantic search while returning an expanded context window (including some neighboring chunks) to the LLM \cite{expanded_context}. This provides a broader context for the model to draw from, increasing the chances of fetching relevant information. 

Re-ranking methods involve an additional step where the initially retrieved documents are sorted again, ensuring that the most relevant pieces of information are prioritized \cite{reranking}. This is typically achieved using cross-encoders, which are capable of scoring pairs of text consisting of the user query and each of the chunks returned by the retriever.

Furthermore, hybrid search techniques combine traditional keyword-based search (such as BM25 algorithm \cite{BM25}) with semantic search capabilities to better capture the context that might be pertinent to a user's query. By leveraging all of these techniques, we were able to noticeably boost the performance of the retriever, and the overall system. It should also be noted that current trends in LLM development include larger contexts which are likely to further improve results.

\subsection{Importance of Metadata}

To enable a question-answering system to effectively query both structured and unstructured data, it is important to ensure that the metadata associated with tables, databases, documents, and other data sources is accurate and well-defined. Metadata, such as column names, data types and relationships, and summary of each table's or document's subject, aids in the correct identification and interpretation of data by the retrieval system and LLMs. Having proper metadata allows a more seamless interaction between LLMs and structured databases by enabling the LLMs to create proper queries using Sequential Query Language (SQL) or other methods as applicable. Conversely, mistakes or ambiguity in SQL code can lead to incorrect or incomplete answers, highlighting the critical role of precise query formulation in building a successful question-answering system using LLMs.

\subsection{Pre-processing of Unstructured Data}

Dealing with unstructured data such as PDF documents and web pages can pose a significant challenge for question-answering systems, as the information is often embedded within the context of the document's layout, such as tables, headings, and paragraphs. In such cases, in addition to providing proper metadata, steps should be taken to preserve the document structure during the conversion to machine-readable text format by recognizing and properly extracting tables, headings, and embedded images or diagrams. This structure should be further preserved during text splitting/chunking process to ensure that e.g., tables are not split across different chunks thereby creating potentially incomplete context. This requires specialized techniques and tools to handle different unstructured data formats effectively, as overlooking this step can result in a loss of valuable information which, in turn, can negatively impact the question-answering system's accuracy.

\subsection{Human-in-the-Loop}

Early prototypes and experiments with LLMs in the spectrum regulation domain will be potentially prone to inaccuracies and bias which in turn can undermine trust in the system and jeopardize the successful integration and adoption of this technology. To address these concerns, having early feedback from human experts can provide a critical layer of oversight. Specifically, domain experts can verify the accuracy of LLM-generated responses, ensure compliance with regulations, and provide valuable feedback to fine-tune the system before fully integrating it into the existing internal or public-facing operations. This human-in-the-loop approach is essential not only for improving the performance and fairness of LLMs but also for upholding the integrity and trustworthiness of regulatory processes in an increasingly AI-driven era. Furthermore, government guidelines may require human participation \cite{automated_decision_tbs}.

\section{Conclusion and Future Prospects}

The use of AI in general, and LLMs in particular, in the spectrum regulatory and policy-making domain holds significant promise for enhancing efficiency of various processes. Automation of routine, yet time-consuming and labour-intensive tasks via LLMs can free up human resources for more strategic and creative work and lead to streamlined administrative processes and cost savings. In this paper we discussed how LLMs can assist in the analysis of regulations and policies, extraction of rules as code, summarizing and interpreting stakeholder positions, finding specific spectrum licence conditions, and other key regulatory workflows. Future work should look at approaches for ensuring transparency and fairness in LLMs in this context as well as potential gains from models fine-tuned on spectrum regulatory domain data.

It should be noted that the human element remains essential to provide oversight, address exceptional cases, and ensure that the LLM agents operate in accordance with regulatory intent and ethical standards. As technology continues to advance, the integration of LLMs into regulatory workflows offers a promising avenue for enhancing the efficiency and effectiveness of radio spectrum regulation while upholding the necessary checks and balances.

\end{document}